\documentclass[twocolumn,pre,tightenlines,superscriptaddress,showpacs]{revtex4-2}

\usepackage{amsmath}
\usepackage{amssymb,amsfonts,latexsym}
\usepackage{bm}
\usepackage[mathcal]{euscript}
\usepackage{graphicx}
\usepackage{epsfig}
\usepackage{color}
\usepackage{xcolor}
\usepackage{psfrag}

\newcommand{\be}{\begin{equation}}
\newcommand{\ee}{\end{equation}}
\newcommand{\bea}{\begin{eqnarray}}
\newcommand{\eea}{\end{eqnarray}}
\newcommand{\ve}{\varepsilon}
\newcommand{\la}{\langle}
\newcommand{\ra}{\rangle}
\newcommand{\nc}{N_{\mathrm{corr}}}

\begin{document}

\title{Nonlinear dielectric response in glasses: restoring forces and avoided spin-glass criticality}

\author{Eric Bertin}
\affiliation{Univ.~Grenoble Alpes, CNRS, LIPhy, 38000 Grenoble, France}

\author{Fran\c{c}ois Ladieu}
\affiliation{SPEC, CEA, CNRS, Universit\'e Paris-Saclay, CEA Saclay, Bat 772, 91191 Gif-sur-Yvette Cedex, France}

\date{\today}
%\pacs{}

\begin{abstract}
Experimental measurements of nonlinear dielectric response in glassformers like supercooled glycerol or propylene carbonate have
been interpreted as providing evidence for a growing thermodynamic length scale when lowering temperature. A heuristic picture based on coherently flipping `superdipoles' with disordered internal structure has been argued to capture the essence of the experimentally reported behavior, pointing to the key role of effectively disordered interactions in structural glasses.
We test these ideas by devising an explicit one-dimensional model of interacting spins incorporating both the spin-glass spirit of the superdipole argument, and the necessary long-time decorrelation of structural disorder, encoded here in a slow dynamics of the coupling constants. The frequency-dependent third-order response of the model qualitatively reproduces the typical humped shape reported in experiments.
The temperature dependence of the maximum value is also qualitatively reproduced. 
In contrast, the humped shape of the third-order response is not reproduced by a simple kinetically constrained spin model with non-interacting spins.
To rationalize these results, we propose a two-length-scale scenario by distinguishing between the characteristic length of dynamical heterogeneities and a coherence length that monitors the effect of interactions. We show that both length scales are identical in the kinetically constrained spin model, while they have significantly different dynamics in the model of interacting spins.
\end{abstract}

\maketitle

\section{Introduction}

Understanding the origin of the very fast increase of the relaxation time of glasses over a moderate temperature range remains a challenging task for condensed matter physics \cite{Berthier2011,Arceri2021}. Experimentally, the glass transition has been studied by a vast range of techniques \cite{Ediger2000}, ranging from Nuclear Magnetic Resonance \cite{Tracht1998} to dielectric spectroscopy \cite{Lunkenheimer2002}, neutron diffraction \cite{Alba2004,Dalle2009}, optical techniques \cite{Cicerone1992,Li1994,Aouadi2000,Wang2000} and even Atomic Force Microscopy \cite{Vidal2000}. As each technique specifically probes some degrees of freedom, comparing the experimental results is of great interest --see e.g. \cite{Richert2011}. This comparison reveals that all of them are well coupled --apart from one exception which is physically understood \cite{Tarjus1995,Richert2011}. Indeed, upon cooling, they exhibit a similar characteristic time scale --called the $\alpha$ relaxation time--, which governs, e.g., both dielectric spectra --probing rotation of molecules-- and viscosity behavior --probing the mechanical response of molecules. Yet, despite the impressive number of experimental studies characterizing the glass transition, the origin of this extremely fast viscous slow down remains debated \cite{Arceri2021,Berthier2011} and opposite theories --see below-- claim to account for the glass phenomenology.

An important insight  has been provided over the last decade by experimental measurements of nonlinear dielectric response in glassformers like supercooled glycerol or propylene carbonate. Motivated by a spin-glass inspired theoretical prediction \cite{Bouchaud2005,Tarzia2010}, these experiments have provided evidence for a growing thermodynamic length scale when lowering temperature \cite{Crauste2010,Brun2012,Bauer2013,Casalini2015,Albert2019}.
In particular, it has been found experimentally that non-linear responses of third \cite{Crauste2010,Brun2012,Bauer2013,Casalini2015} and fifth order \cite{Albert2016} have a characteristic humped shape when plotted as a function of frequency, and that their maximum value, that grows when temperature is lowered --or pressure increased \cite{Casalini2015}-- is a measure of a coherence length scale of the glass.
Such a growing coherence length, together with Arrhenius law, could account for such a dramatic increase of relaxation time provided the free energy barrier grows typically as a power law of the coherence length.
From these measurements, a heuristic picture inspired by spin-glass physics has emerged \cite{Ladieu2012,Buchenau2017}, namely the idea that global relaxation results from the coherent relaxation of `superdipoles' having a size approximately equal to the coherence length. Quite importantly, these superdipoles have a disordered and essentially frozen internal structure in terms of the microscopic dipoles, and their coherent relaxation is key to rationalize the experimentally observed behavior of the non-linear response as a function of frequency and temperature \cite{Gadige2017,Albert2019}, including works done previously in another perspective \cite{Richert2006,Young2017,Richert2017}. It has also been argued \cite{Albert2016,Biroli2021} that such a coherent relaxation of superdipoles of increasing size (when temperature is lowered) implies that thermodynamic aspects play an important role in the glassy relaxation, thereby a priori ruling out the facilitation scenario put forward in kinetically constrained models, at least in their simplest implementations.
However, no consensus has been reached yet in the glass community regarding this issue \cite{Speck2019,Speck2021},
and recent extensive numerical simulations of low temperature glasses indicate that dynamic facilitation could still play a major role in this regime
\cite{Guiselin2022,Scalliet2022}.

This lack of consensus is partly due to the fact that nonlinear responses are notoriously difficult to evaluate, up to the point that explicit calculations are possible only in oversimplified frameworks, such as phenomenological models \cite{Ladieu2012,Buchenau2017}, or frameworks where the
mechanism leading to a supercooled state plays no direct role  \cite{Diezemann2012,Diezemann2013,Diezemann2018,Diezemann2022}. Moving to the theories accounting for the glass transition in itself, whatever the viewpoint adopted about the importance of thermodynamics aspects, the proposed theoretical reasonings mostly rely on general and plausible arguments \cite{Gadige2017,Biroli2021,Speck2021} and not on fully explicit calculations. Although quite successful in terms of comparison with the experimental data \cite{Crauste2010,Albert2016,Speck2021}, these general arguments leave a number of important questions open, such as: Which type of interactions between microscopic dipoles could generate an emerging phenomenology in terms of superdipoles with disordered but coherent internal structure?
How can these superdipoles melt in the long-time regime to recover a trivial non-linear response at very low frequency, and thereby reproduce the humped shape of non-linear responses (in particular the third order one)?

In this paper, we propose two explicit one-dimensional models, which both cannot have any long-range thermodynamic order at finite temperature. Our two models differ only about the importance of interactions between effective degrees of freedom, and we compare their nonlinear responses in frequency and temperature. On one side we devise an explicit one-dimensional model of interacting spins incorporating both the spin-glass spirit of the superdipole argument, and the necessary long-time decorrelation of structural disorder, encoded here in a slow dynamics of the coupling constants between neighboring spins. We find that the frequency-dependent third-order response of the model qualitatively reproduces the typical humped shape reported in experiments, and that the static third-order response is that of non-interacting spins.
The temperature dependence of the maximum value is also qualitatively reproduced. We compare step by step these results with those obtained with our second model, which is a simple kinetically constrained spin model inspired by the Fredrickson-Andersen model. We find that the cubic response of the kinetically constrained model monotonically decreases as a function of frequency.
These results are rationalized through a two-length scale scenario, which distinguishes between the characteristic length of dynamical heterogeneities and a coherence length that monitors the effect of interactions. Our results indicate that both length scales are identical in the kinetically constrained spin model, while they have significantly different dynamics in the model of interacting spins. These results provide some quantitative evidence in favor of the landscape-driven restoring force in systems of interacting spins with slowly evolving coupling constants.

\section{Stochastic spin models}
\label{sec:models}

\subsection{Disordered spin model with slowly evolving couplings}
\label{sec:spin:interact}

\subsubsection{Spin dynamics}

We consider a one-dimensional spin model with $N$ spins $S_i=\pm 1$ ($i=1,\dots,N$), with periodic boundary conditions ($S_{N+1}\equiv S_1$). An external field $E(t)$, playing the role of the electric field in the experiment, is applied.
Neighboring spins $S_i$ and $S_{i+1}$ on the lattice interact via a link-dependent coupling constant $J_{i,i+1}$.
The time-dependent Hamiltonian of the model reads
\be \label{eq:hamilton}
H(t) = -\sum_{i=1}^N J_{i,i+1} S_i S_{i+1} - E(t) \sum_{i=1}^N S_i.
\ee
The stochastic dynamics is constrained by the detailed balance property, valid for a static field $E(t)=E_0$.
More specifically, spins are assumed to obey a stochastic reversal dynamics satisfying detailed balance with respect to the equilibrium distribution associated with the Hamiltonian $H$ (corresponding to a static field $E_0$),
$P_{\rm eq} \propto e^{-\beta H}$, where $\beta=1/k_B T$ is the inverse temperature.
The probability per unit time to flip a spin is chosen according to the Glauber rate
\be \label{eq:rate:S}
W(-S_i|S_i) =  \frac{\nu_0}{1+e^{\beta \Delta H_i^S}}
\ee
where $\Delta H_i^S$ is the energy change induced by the reversal of the spin $S_i$,
and where $\nu_0$ is the characteristic attempt frequency of the spin dynamics.
We further assume that the definition \eqref{eq:rate:S} of the transition rates remains valid for a time-dependent field $E(t)$, leading to time-dependent transition rates: this is well justified --see e.g.~Chap.~$14$ of Ref.~\cite{kremer2002}-- in our case where the frequency of the field is much smaller than $\nu_0$.

\subsubsection{Coupling dynamics}

In the form \eqref{eq:hamilton}, the Hamiltonian is very similar to that of a spin-glass model.
This spin-glass-like form of the Hamiltonian is motivated by the standard heuristic argument describing a dielectric glass as a set of superdipoles, each one being made of frozen and disordered arrangement of electric dipoles \cite{Ladieu2012} (see also Sec.~\ref{sec:superdipoles}).
This phenomenological argument, deeply rooted in the spin-glass physics, predicts a divergence of the non-linear dielectric responses when a coherence length,
characterizing the size of superdipoles, increases. The argument also correctly predicts the absence of divergence for the linear dielectric response.

In spite of this qualitative success, one of the difficulties with the above heuristic argument is that it does not describe how the nonlinear dielectric response becomes small again at very low frequencies. Intuitively, this is the regime where the disorder inside the superdipoles unfreezes due to, e.g., the fact that molecules are anisotropic objects which mutual interaction depends onto their relative orientation --for example the dipole-dipole interaction changes its sign depending on the angle between two molecular dipoles.
Therefore as the system relaxes, the mutual orientations of molecules changes and their interactions are modified and may even change sign.
Assuming that couplings $J_{i,i+1}$ in our model play a role similar to the interactions between molecules --which is implicit in Eq.~(\ref{eq:hamilton})-- and are thus key to the glass transition, the couplings $J_{i,i+1}$ would also be expected to change their sign in the long run.
To include this effect explicitly in the model, we assume that the coupling constants $J_{i,i+1}$ are not completely frozen but have a slow dynamics, on a time scale much longer than the time scale of the spin dynamics (see also \cite{Penney1993} for a closely related model in the context of neural networks).
For simplicity, we choose bivalued coupling constants $J_{i,i+1}=\pm J_0$, and assume a stochastic reversal dynamics
$J_{i,i+1} \to -J_{i,i+1}$ with a transition rate satisfying detailed balance with respect to the equilibrium distribution 
$P_{\rm eq}$.
The probability per unit time to reverse the sign of the coupling constant $J_{i,i+1}$ is assumed to be
\be \label{eq:rate:J}
W(-J_{i,i+1}|J_{i,i+1}) = \frac{\nu_1}{1+e^{\beta \Delta H_{i,i+1}^J}}
\ee
where $\Delta H_{i,i+1}^J$ is the energy change induced by the reversal of the coupling constant $J_{i,i+1}$,
and where $\nu_1$ is the characteristic attempt frequency of the dynamics.
As we expect coupling constants to evolve on much larger time scales than the spins, we assume that $\nu_1 \ll \nu_0$, so that the coupling constants appear as essentially frozen on the time scale of the spin dynamics.

In the following, we assume that the characteristic frequency $\nu_1$ depends on temperature according to an Arrhenius law,
\be \label{eq:nu1:arrhenius}
\nu_1(T) = \nu_0\, e^{-B/T},
\ee
where $B$ is a typical energy barrier for rearrangements. This choice takes into account, at a qualitative level, the thermally activated nature of rearrangements, thereby inducing at low temperature $T$ a time-scale separation between the fast spin dynamics and the slow coupling dynamics.

\subsection{Kinetically constrained model}
\label{sec-def-kcm}

We wish to compare the above model of interacting spins with stochastic couplings to a simple kinetically constrained model (KCM).
As a minimal KCM, we consider a simple extension of the Fredrickson-Andersen (FA) model \cite{Ritort2003}, that includes spin variables on top of the usual mobility excitations. KCM with two local variables have been previously considered in other contexts, like ion mobility in glasses \cite{Schulz1998}.
In the usual FA model, mobility excitations are the only degrees of freedom (so that kinetic constraints only affect mobility excitations themselves), while in our model mobility excitations may couple to other physical degrees of freedom. 
More explicitly, we introduce local facilitation variables $n_i=0$ or $1$ on each site $i$, where $n_i=1$ corresponds to the presence of a mobility excitation on site $i$.
As in the FA model, mobility excitations are assumed to be non-interacting, and thus to contribute to the Hamiltonian through a term proportional to $\sum_i n_i$.
The generalized Hamiltonian then reads
\be \label{eq:hamilton:kcm}
\tilde{H} = - E(t) \sum_{i=1}^N S_i + K \sum_{i=1}^N n_i \,,
\ee
with the external field $E(t)$ and a characteristic energy $K$ of mobility excitations.
A spin $S_i$ can be flipped only when $n_i=1$, which leads to a slowdown of the dynamics at low temperature, because mobility excitations become rare due to their energetic cost. The transition rate for spin reversal reads:
\be \label{eq:rate:S:kcm}
W(-S_i|S_i) = \frac{\nu_0 \, n_i}{1+e^{\beta \Delta \tilde{H}_i^S}}
\ee
where $\Delta \tilde{H}_i^S=2ES_i$ is the variation of the Hamiltonian $\tilde{H}$ defined in Eq.~\eqref{eq:hamilton:kcm} associated with the transition
$S_i \to -S_i$. Here again, we assume the transition rates to be slowly time-dependent due to the field $E(t)$.
As for the dynamics of mobility excitations, we follow the standard rules of the FA model.
The local variable $n_i$ can only change its value if at least one of the neighboring variables
$n_{i-1}$ or $n_{i+1}$ is equal to $1$. On a coarse-grained scale, this kinetic constraint on the dynamics of the variables $n_i$ leads to an effective diffusion of mobility excitations \cite{Ritort2003}.
To fulfill detailed balance with respect to the Hamiltonian (\ref{eq:hamilton:kcm}),
and to take into account kinetic constraints on the dynamics  of mobility excitations,
we choose the following form for the transition rate from $n_i$ to $n_i'=1-n_i$,
\be \label{eq:rate:n:kcm}
W(1-n_i|n_i) = \frac{\nu_0}{1+e^{\beta \Delta \tilde{H}_i^n}}\,\theta(n_{i-1}+n_{i+1})\,,
\ee
where $\theta(x)$ is the Heaviside function, $\theta(x)=1$ if $x>0$ and $\theta(x)=0$ otherwise; $\Delta \tilde{H}_i^n = K(1-2n_i)$ is the variation of the Hamiltonian $\tilde{H}$ defined in Eq.~\eqref{eq:hamilton:kcm} associated with the transition $n_i \to 1-n_i$.

\section{Static third-order response}
\label{sec:static:resp}

\subsection{General expression of cubic responses}

We first consider a generic spin model at equilibrium, with a Hamiltonian $H$, which is a function of $N$ spin variables $\{S_i\}$ and possibly of other variables present in the system.
Spins are coupled to a static external field $E_0$, so that the Hamiltonian takes the form
\be
H = H_0 - E_0 \sum_i S_i
\ee
where $H_0$ is the Hamiltonian in the absence of external field ($E_0=0$).
We assume that $H_0$ is invariant by global spin reversal $\{S_i\}\to\{-S_i\}$.
The free energy density is defined by
\be \label{eq:def:free:ener}
f = -\frac{1}{\beta N} \, \ln Z
\ee
where $Z=\sum_{\mathcal{C}} e^{-\beta H(\mathcal{C})}$ is the partition function; $\mathcal{C}$ is a short-hand notation for the list of all microscopic variables, including the $N$ spins $S_i$.
The average magnetization $\la m \ra$, where $m=\frac{1}{N}\sum_{i=1}^N S_i$, is given by
\be \label{eq:m:f:deriv}
\la m \ra = -\frac{\partial f}{\partial E_0}\,.
\ee
Static linear and non-linear responses are obtained by expanding $\la m\ra$ for small $E_0$,
\be \label{eq:m:expansion}
\la m \ra = \chi_1^s E_0 + \chi_3^s E_0^3 + \dots
\ee
where we have kept terms only up to third order, and used the spin-reversal symmetry to eliminate even terms in $E_0$ in the expansion. This leads in particular to a definition of a static cubic response as
\be \label{eq:chi3s}
\chi_3^s = \frac{1}{6} \, \frac{\partial^3 \la m \ra}{\partial E_0^3}\,.
\ee
An alternative expression of the cubic response is obtained by considering a field $E_0+\ve$ and evaluating the linear response to the tiny contribution $\ve \ll E_0$, in the presence of the small field $E_0$. Here, one considers the linear response
\be \label{eq:def:chislin}
\chi_{\rm lin}^s(E_0) = \frac{\partial \la m\ra}{\partial E_0}
\ee
and expands it to quadratic order in $E_0$,
\be
\chi_{\rm lin}^s(E_0) = \chi_1^s + \chi_{21}^s E_0^2 + \dots,
\ee
which defines the cubic response $\chi_{21}^s$.
Using Eqs.~(\ref{eq:m:expansion}) and (\ref{eq:def:chislin}), we obtain
\be
\chi_{\rm lin}^s = \chi_1^s + 3\chi_3^s E_0^2 + \dots
\ee
%The quadratic contribution is equal to $\chi_{21}^s E_0^2$, by definition of the cubic response $\chi_{21}^s$.
One thus has the simple relation between the cubic responses $\chi_3^s$ and $\chi_{21}^s$:
\be \label{eq:chi21:chi3}
\chi_{21}^s = 3\chi_3^s\,.
\ee
As we will see below in Sec.~\ref{sec:dynamic:resp},
the static response $\chi_{21}^s$ corresponds to the zero-frequency limit of the dynamic cubic response considered in this paper, and we thus focus on $\chi_{21}^s$ rather than on $\chi_3^s$ in the following.

From Eqs.~(\ref{eq:chi3s}) and (\ref{eq:chi21:chi3}) we have
\be \label{eq:chi21s}
\chi_{21}^s = \frac{1}{2} \, \frac{\partial^3 \la m \ra}{\partial E_0^3}\,.
\ee
Using Eq.~(\ref{eq:m:f:deriv}), we eventually end up with
\be \label{eq:chi21static}
\chi_{21}^s = -\frac{1}{2} \frac{\partial^4 f}{\partial E_0^4}\,.
\ee
This general expression of the static third-order response $\chi_{21}^s$ can now be applied to the two spin models introduced in Sec.~\ref{sec:models}.

\subsection{Spin model with stochastic couplings}

We consider the spin model with stochastic couplings defined in Sec.~\ref{sec:spin:interact}, with a static external field $E(t)=E_0$.
Using the expression (\ref{eq:hamilton}) of the Hamiltonian $H$, the partition function reads
\be
Z = \sum_{\{S_i\},\{J_{i,i+1}\}} e^{\beta \sum_{i=1}^N J_{i,i+1} S_i S_{i+1} + \beta E_0 \sum_{i=1}^N S_i}
\ee
and it can be determined for instance using a standard transfer matrix technique. However, a simpler calculation can be performed using a change of summation variable. Defining $\sigma_i=J_{i,i+1} S_i S_{i+1}/J_0$, the partition function takes the simpler form
\be
Z = \sum_{\{S_i\},\{\sigma_i\}} e^{\beta J_0 \sum_{i=1}^N \sigma_i + \beta E_0 \sum_{i=1}^N S_i}
\ee
which now effectively involves only non-interacting degrees of freedom
(the sum is performed over all values $S_i=\pm 1$ and $\sigma_i=\pm 1$ for $i=1,\dots,N$).
One thus finds
\be
Z = \big[ 4 (\cosh \beta J_0) \cosh(\beta E_0) \big]^N.
\ee
The free energy density defined in Eq.~(\ref{eq:def:free:ener}) then reads
\be \label{eq:free:ener:sc}
f(\beta,E_0) = f_{\rm int}(\beta) + f_{\rm id}(\beta,E_0)\,,
\ee
where
\be
f_{\rm int}(\beta) = -\frac{1}{\beta} \ln [2 \cosh(\beta J_0)]
\ee
is the contribution to the free energy density resulting from (annealed) random interactions between spins, and
\be \label{eq:id:free:ener}
f_{\rm id}(\beta,E_0) = -\frac{1}{\beta} \ln [2\cosh(\beta E_0)]
\ee
is the free energy density of the ideal spin gas.
Hence at static level, interactions between spins are decoupled from the external field, in the sense that their respective contribution to the free energy density are additive.
Eqs.~\eqref{eq:m:f:deriv}, \eqref{eq:m:expansion} and \eqref{eq:chi21static} then lead to simple expressions for the static linear and third-order responses respectively,
\be \label{eq:chi21static:intspins}
\chi_1^s = \beta, \qquad \chi_{21}^s = -\beta^3
\ee
which are nothing but the linear and third order responses of an ideal spin gas, i.e., a paramagnetic system of noninteracting spins.
The linear response exhibits a moderate increase, $\propto 1/T$, when lowering temperature, in qualitative agreement with experiments.
Note that only $f_{\rm id}(\beta,E_0)$ contributes to the response $\chi_{21}^s$,
because only this contribution to the free energy density depends on the field $E_0$.

\subsection{Kinetically contrained spin model}

For the kinetically constrained spin model introduced in Sec.~\ref{sec-def-kcm},
considered here with a static field $E(t)=E_0$, the partition function defined by the Hamiltonian $\tilde{H}$ given in Eq.~(\ref{eq:hamilton:kcm}) reads as
\be
Z_{\rm kc} = \sum_{\{S_i\},\{n_i\}} e^{\beta E_0 \sum_{i=1}^N S_i-\beta K \sum_{i=1}^N n_i}.
\ee
The spins $S_i$ and mobility excitations $n_i$ are non-interacting variables in the Hamiltonian $\tilde{H}$, so that the partition function simply factorizes as
\be
Z_{\rm kc} = \big[ 4 (\cosh \beta K) \cosh(\beta E_0) \big]^N.
\ee
The free energy density again takes an additive form
\be
f_{\rm kc}(\beta,E_0) = f_{\rm mob}(\beta) + f_{\rm id}(\beta,E_0)\,,
\ee
where
\be
f_{\rm mob}(\beta) = -\frac{1}{\beta} \ln [2 \cosh(\beta K)]
\ee
is the free energy contribution of mobility excitations, and where the ideal spin gas contribution $f_{\rm id}(\beta,E_0)$ has the same expression as in Eq.~(\ref{eq:id:free:ener}).
Hence, one also finds for the kinetically constrained model that the linear and third-order static responses are given by the ideal spin gas responses
$\chi_1^s=\beta$ and $\chi_{21}^s = -\beta^3$ respectively,
as in Eq.~(\ref{eq:chi21static:intspins}).
Note that this result was expected since spins are non-interacting in the present model.

\section{Dynamic third-order response}
\label{sec:dynamic:resp}

Our goal is to evaluate the third-order (dielectric) response of the polarization (i.e., magnetization in the spin language) to a time-dependent external field $E(t)$ oscillating at (angular) frequency $\omega$. 
Here again, there are several ways to define a third-order response. For instance, one may consider either the response to the field at frequency $\omega$ or at frequency $3\omega$. The response at frequency $\omega$ can itself be divided into two distinct response functions. As in the static case, we focus here on the simplest third-order response function,
called $\chi_{21}(\omega)$, which consists in looking at the third-order response at frequency $\omega$ when applying a field $E(t)=E_0+\ve\cos(\omega t)$, in the limit where both the static component $E_0$ and the amplitude $\ve$ are small, with the further assumption that $\ve \ll E_0$.
The third-order response $\chi_{21}(\omega)$, which has been measured experimentally \cite{Lhote2014}, is linear in $\ve$ and quadratic in $E_0$, yielding an overall third-order response in the field amplitude.
It has been shown that all types of third-order responses behave in a similar way \cite{Gadige2017,Lhote2014}, and it is thus legitimate to focus on a specific type of response.

\subsection{Fluctuation-dissipation relation}

The advantage of the third-order response function $\chi_{21}(\omega)$
is that it consists in a linear response to the oscillating contribution of the field. In other words, it is the correction at order $E_0^2$ to the linear response $\chi(\omega,E_0)$ of the polarization at frequency $\omega$,
\be \label{eq:def:chi21dyn}
\chi_{21}(\omega) = \frac{1}{2} \, \frac{\partial^2 \chi}{\partial E_0^2}(\omega,E_0=0)\,.
\ee
Interestingly, the linear response $\chi(\omega,E_0)$ in the presence of a static field $E_0$ can be expressed in terms of the equilibrium correlation function of the magnetization using the fluctuation-dissipation theorem (FDT) since we are dealing with a close-to-equilibrium situation.
We first formulate the FDT in the time domain before moving to the frequency domain. Let us define the (normalized) equilibrium two-time correlation function of the magnetization, $m(t)=N^{-1}\sum_{i=1}^N S_i(t)$ as
\be \label{eq:def:correl}
C(t,E_0) = N \big( \la m(t) m(0) \ra_{E_0} -  \la m \ra_{E_0}^2 \big) ,
\ee
where the notation $\la \dots \ra_{E_0}$ indicates an average over the equilibrium dynamics under a static field $E_0$. The FDT then reads
\be \label{eq:FDT:time}
\chi(t,E_0) = -\beta \theta(t) \, \frac{\partial C}{\partial t}(t,E_0)
\ee
where $\chi(t,E_0)$ is the impulse response [i.e., the linear response of $\la m(t) \ra_{E_0}$ to a Dirac delta in $\ve(t)$ for a field $E(t)=E_0+\ve(t)$], and $\theta(t)$ is the Heaviside function which accounts for the causality condition.

In practice, we may thus use the following procedure to determine numerically the third-order response $\chi_{21}(\omega)$.
One first determines the equilibrium spin correlation function
$C(t,E_0)$ in the time domain for different small values of the static field $E_0$, and then take its Fourier-Laplace transform $\hat{C}(\omega,E_0)$, defined as
\be \label{eq:def:Comega}
\hat{C}(\omega,E_0) = \int_0^{\infty} dt \, e^{i\omega t} \, C(t,E_0)\,.
\ee
In Fourier space, the FDT \eqref{eq:FDT:time} reads
\be \label{eq:FDT}
\chi(\omega,E_0) = \beta C(t=0,E_0) + \beta i\omega\, \hat{C}(\omega,E_0)\,.
\ee
%where $\chi(\omega,E_0)$ is the linear response of the polarization at frequency $\omega$, in the presence of a static field $E_0$.
Note that we consider here both the real and imaginary parts of the Fourier transform of the fluctuation-dissipation relation (\ref{eq:FDT:time}), while standard forms of the fluctuation-dissipation relation in Fourier space usually include only the imaginary part of Eq.~(\ref{eq:FDT}), corresponding to the loss modulus.

Setting $E_0=0$ in Eq.~(\ref{eq:FDT}), one gets the linear response function $\chi_{1}(\omega)$,
\be \label{eq:chi1:cor}
\chi_1(\omega) = \beta C(0,0) + \beta i\omega\, \hat{C}(\omega,0)\,.
\ee
The third-order response function $\chi_{21}(\omega)$ is obtained by
applying the definition (\ref{eq:def:chi21dyn}) to the fluctuation-dissipation relation \eqref{eq:FDT}, yielding
\be \label{eq:chi21:cor}
\chi_{21}(\omega) =  \frac{\beta}{2} \frac{\partial^2 C}{\partial E_0^2}(0, 0) +
\frac{1}{2}\beta i\omega \, \frac{\partial^2 \hat{C}}{\partial E_0^2}(\omega,0)\,.
\ee
Note that in practice, one needs to determine numerically the correlation function with high accuracy in order to evaluate the second derivative of the correlation function with respect to $E_0$.

\begin{figure}[t]
%  \centering
  \includegraphics[width=0.48\textwidth]{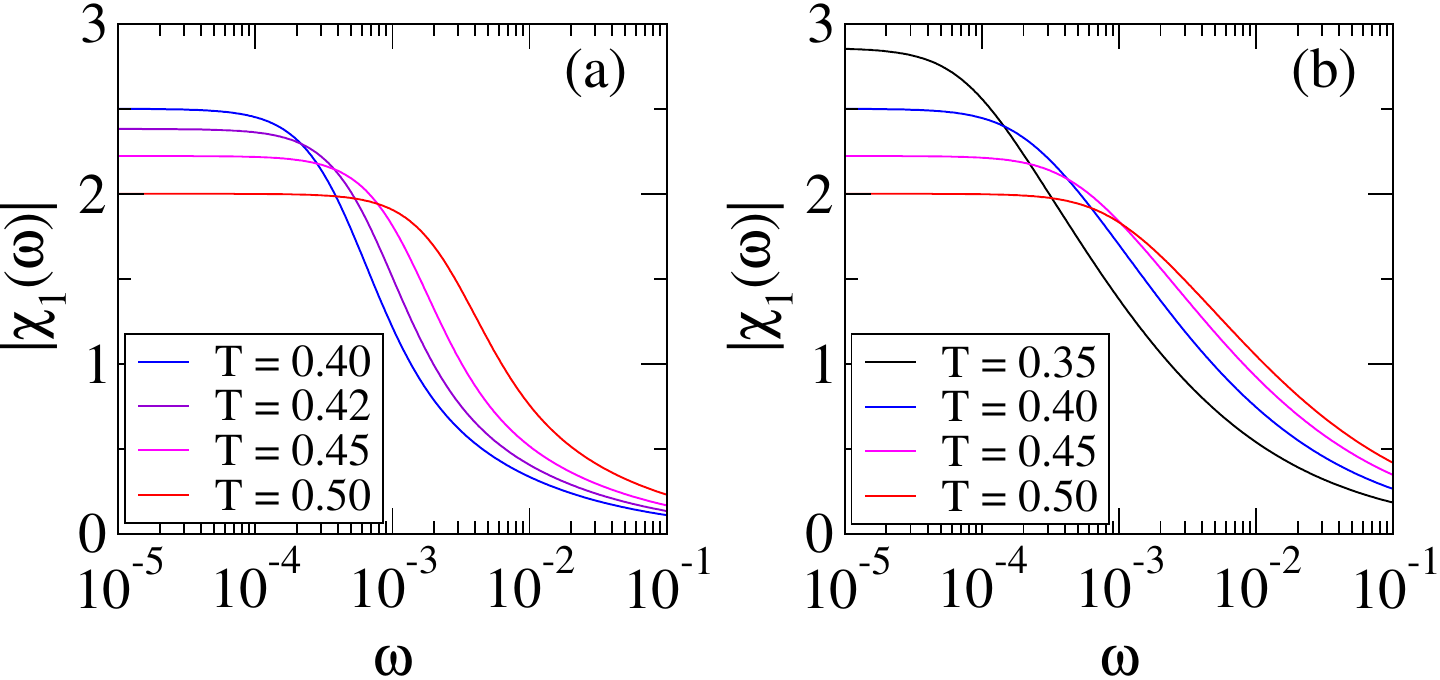}
  \caption{Modulus of the linear response $\chi_{1}(\omega)$ as a function of the angular frequency $\omega$, for different values of temperature (same color code on both panels). (a) Spin model with stochastic couplings for $T=0.40$, $0.42$, $0.45$, $0.50$, from top to bottom at low $\omega$ ($J_0=1$, $B=3$).
  (b) Kinetically constrained spin model for $T=0.35$, $0.40$, $0.45$, $0.50$, from top to bottom at low $\omega$ ($K=1$).
  In both cases, the response function decays monotonously with frequency. System size: $N=10^3$ for (a) and (b).}
  \label{fig:chilin}
\end{figure}

\subsection{Spin model with stochastic couplings}
\label{sec:nonlinresp:spin:couplings}

\begin{figure}[t]
%  \centering
  \includegraphics[width=0.48\textwidth]{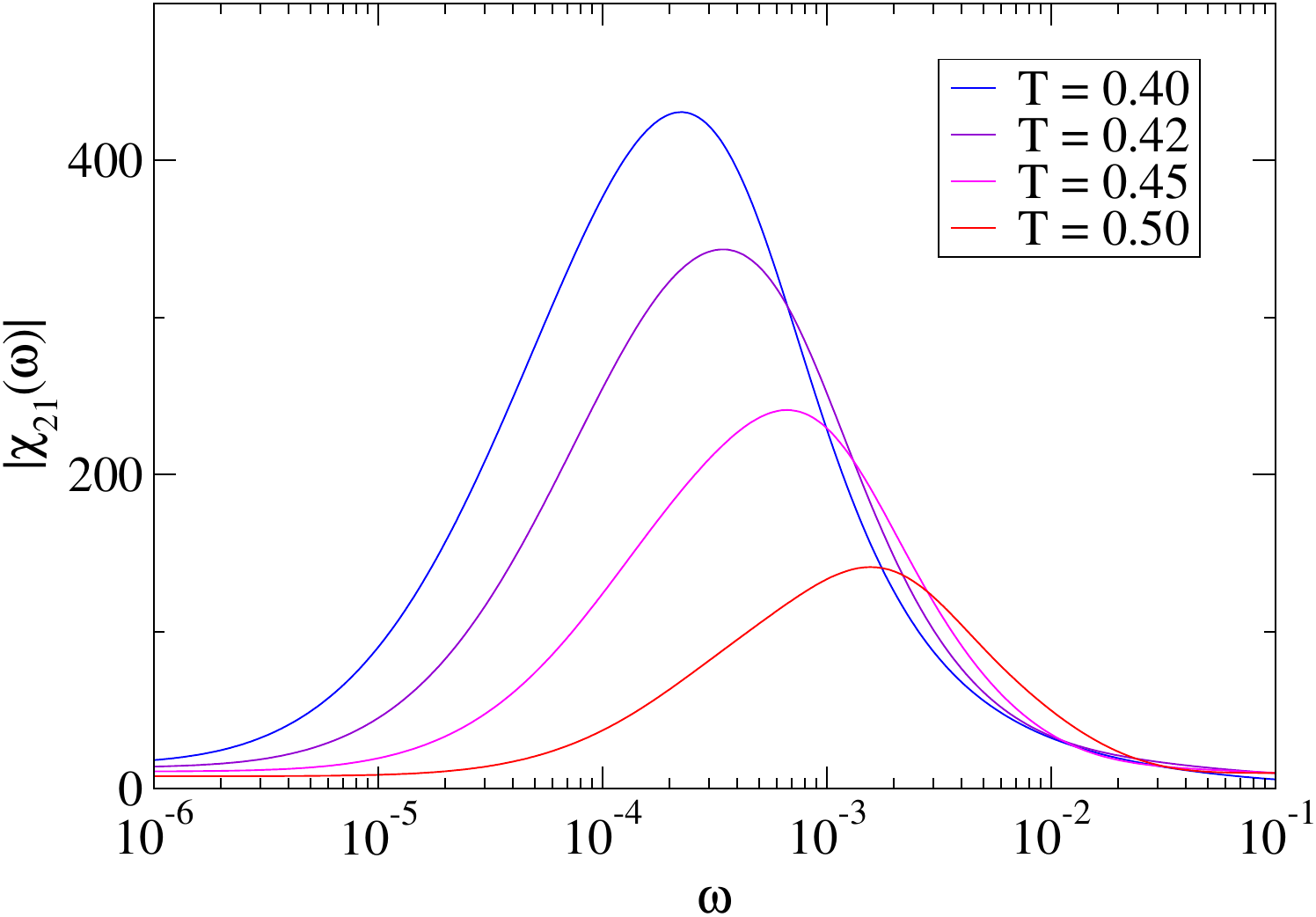}
  \caption{Modulus of the third-order response $\chi_{21}(\omega)$ for different values of temperature ($T=0.40$, $0.42$, $0.45$, $0.50$, from top to bottom) in the spin model with stochastic couplings, exhibiting a pronounced peak whose height increases when decreasing temperature, while peak frequency decreases with temperature. Parameters: $J_0=1$, $B=3$, $N=10^3$. The third-order response is obtained from the time correlation $C(t,E_0)-C(t,0)$ evaluated for $E_0=0.04$.}
  \label{fig:chi21}
 \end{figure}

We have determined numerically both the linear and the third-order response functions $\chi_1(\omega)$ and $\chi_{21}(\omega)$
using kinetic Monte-Carlo simulations of the spin model defined by Eqs.~(\ref{eq:hamilton}), (\ref{eq:rate:S}) and (\ref{eq:rate:J}).
To obtain the frequency-dependent response functions $\chi_1(\omega)$ and $\chi_{21}(\omega)$ over a broad range of frequencies, we first perform accurate fits of the time-dependent correlation function $C(t,E_0)$ for $E_0=0$ and for a small, non-zero value $E_0$.
Fits of $C(t,E_0)$ are constrained to take the known equilibrium value
\be
C(0,E_0) = N\big(\langle m^2 \rangle_{E_0}-\langle m \rangle_{E_0}^2\big)\,,
\ee
which is computed from the second derivative of the free energy (\ref{eq:free:ener:sc}) with respect to the field.
The linear response function is evaluated from $C(t,0)$ using Eq.~(\ref{eq:chi1:cor}). The third-order response is obtained from
Eq.~(\ref{eq:chi21:cor}), using the approximation
\be
\frac{\partial^2 C}{\partial E_0^2}(t,0) \approx \frac{2}{E_0^2}\,[C(t,E_0)-C(t,0)]\,,
\ee
which holds since $C(t,E_0)$ is an even function of $E_0$.
The modulus $|\chi_1(\omega)|$ of the linear response is plotted on Fig.~\ref{fig:chilin}(a) for several temperature values.
The modulus $|\chi_{21}(\omega)|$ of the third-order response is plotted on Fig.~\ref{fig:chi21} for different values of the temperature $T$ (we set $k_B=1$).
At a qualitative level, the response function is seen to have the typical humped shape reported in experiments.

Simulations have been performed using a moderate system size $N=10^3$, and averaging over $10^5$ independent runs, to get accurate data.
This procedure has been found to yield cleaner data than simulations of a larger system averaged over a lower number of runs.
We checked that the system size considered remains much larger than the coherence length (see Sec.~\ref{sec:lengths}).

\begin{figure}[t]
%  \centering
  \includegraphics[width=0.48\textwidth]{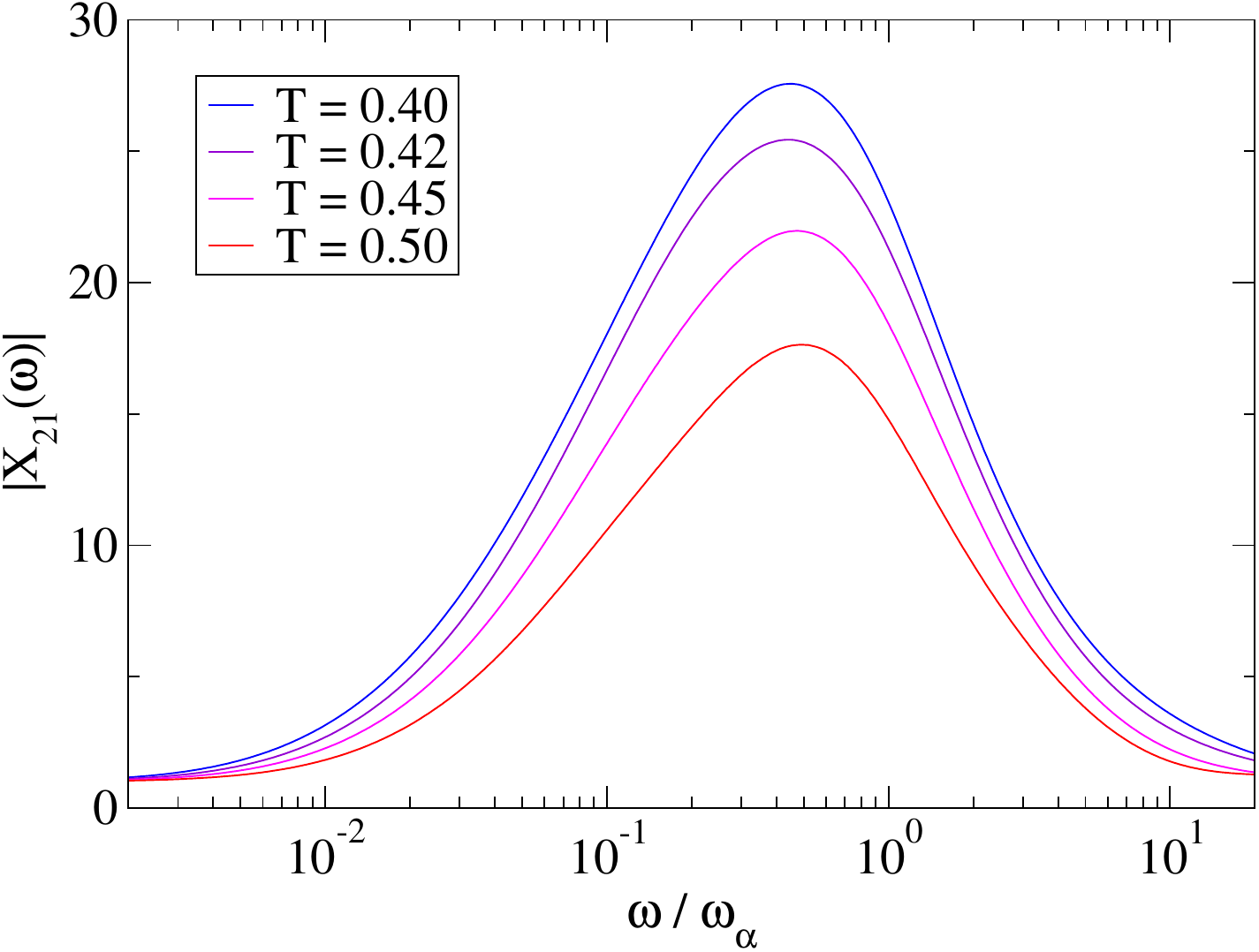}
  \caption{Modulus $|X_{21}(\omega)|$ of the rescaled third-order response as a function of the rescaled frequency $\omega/\omega_{\alpha}(T)$, for different values of temperature ($T=0.40$, $0.42$, $0.45$, $0.50$, from top to bottom; same data as Fig.~\ref{fig:chi21}). The peak value increases when decreasing temperature.}
  \label{fig:X21}
\end{figure}

To investigate the effect of temperature, we define a rescaled third-order response $X_{21}(\omega)=T^3 \chi_{21}(\omega)$ that normalizes the response $\chi_{21}(\omega)$ by the static third-order response of non-interacting dipoles, which is equal to $1/T^3$ in the present model [see Eq.~\eqref{eq:chi21static:intspins}] --or proportional to $1/T^3$ in experiments. Any temperature dependence of the curve $X_{21}(\omega)$ is thus expected to be due to interactions.
We have plotted $X_{21}(\omega)$ in Fig.~\ref{fig:X21} as a function of the rescaled frequency $\omega/\omega_{\alpha}(T)$, where $\omega_{\alpha}(T)$ is the value of $\omega$ for which the loss modulus $\chi_1''(\omega)$, that is the imaginary part of the linear response function, is maximal ($\tau_{\alpha}=2\pi/\omega_{\alpha}$ is the relaxation time).
We observe that in this rescaled representation, the peak value still increases when decreasing temperature, in qualitative agreement with experimental results \cite{Crauste2010}.
Note that to obtain these results, one needs to take into account an increased time scale separation between spin and coupling dynamics when temperature is lowered, as accounted for by the Arrhenius law in Eq.~(\ref{eq:nu1:arrhenius}).

\subsection{Kinetically constrained spin model}

\begin{figure}[t]
%  \centering
  \includegraphics[width=0.48\textwidth]{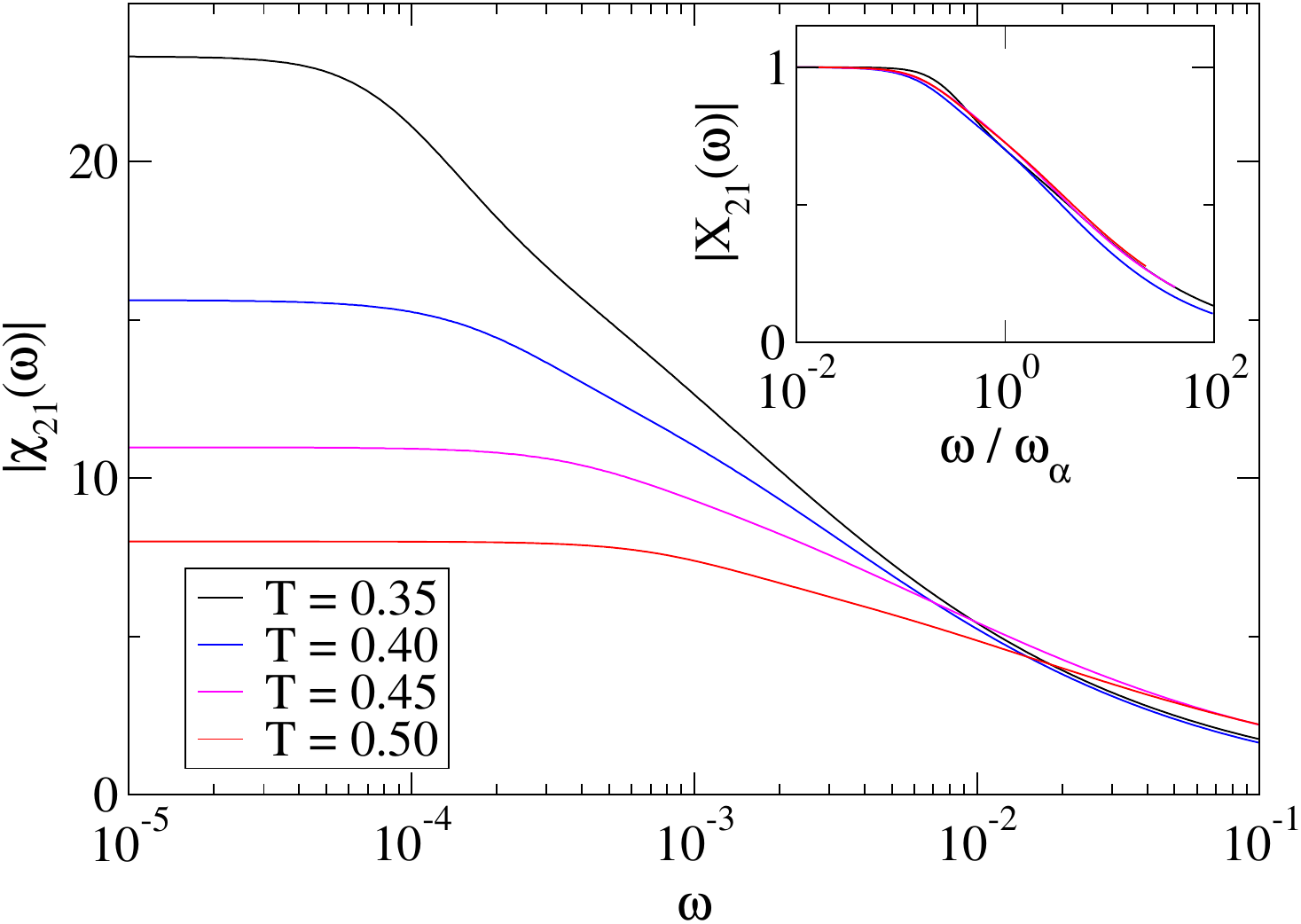}
  \caption{Modulus $|\chi_{21}(\omega)|$ of the third-order response for different values of temperature ($T=0.35$, $0.40$, $0.45$, $0.50$, from top to bottom) in the kinetically constrained spin model. No peak is observed, and $|\chi_{21}(\omega)|$ decreases monotonically with frequency. Inset: corresponding rescaled response $|X_{21}(\omega)|$ versus rescaled frequency $\omega/\omega_\alpha(T)$. Parameters: $K=1$, $N=10^3$.
 The third-order response is obtained from $C(t,E_0)-C(t,0)$ evaluated for $E_0=0.13$, $0.15$, $0.18$, $0.20$ at temperature $T=0.35$, $0.40$, $0.45$, $0.50$, respectively.}
  \label{fig:chi21:KCM}
 \end{figure}

We have also evaluated the dynamic linear and cubic responses $\chi_1(\omega)$
and $\chi_{21}(\omega)$ in the kinetically constrained spin model defined in Sec.~\ref{sec-def-kcm}.
The same fitting protocol as the one described in Sec.~\ref{sec:nonlinresp:spin:couplings} is used.
The modulus $|\chi_1(\omega)|$ of the linear response is plotted in Fig.~\ref{fig:chilin}(b), and is observed to monotonously decrease with frequency as expected.
The modulus $|\chi_{21}(\omega)|$ of the cubic response is plotted in Fig.~\ref{fig:chi21:KCM}. We see that contrary to the model of interacting spins, no peak is observed and $|\chi_{21}(\omega)|$ decreases monotonically as function of the frequency $\omega$.
The curves approximately collapse to a master curve when rescaled by $1/T^3$, up to a simultaneous rescaling of frequency into $\omega/\omega_{\alpha}(T)$.
The corresponding plot of the rescaled response $|X_{21}(\omega)|=T^3 |\chi_{21}(\omega)|$ versus 
$\omega/\omega_{\alpha}(T)$ is displayed in the inset of Fig.~\ref{fig:chi21:KCM}.

\section{A two-length-scale scenario}
\label{sec:lengths}

Non-linear dielectric responses are mostly used as experimental tools to probe the presence of a coherence length in glasses.
Numerically, more direct measures of a coherence length can also be performed. We argue below in favor of a two-length scale scenario in the dynamics of the glassy spin model with stochastic couplings.
This scenario allows us to evidence the key role played by interactions to generate a dynamic coherence length that becomes significantly larger than the dynamic length scale characterizing dynamical heterogeneities.
We start by recalling the heuristic `superdipole' argument, that is useful to grasp the physical picture behind the humped shape of the cubic response.

\subsection{Superdipole picture}
\label{sec:superdipoles}

\subsubsection{Physical motivation}

Obtaining the humped shape of the cubic response is non trivial in the sense that it requires two important ingredients of the model, namely the presence of interactions between spins and the time scale separation between the dynamics of the spins and that of the couplings.
In the absence of interactions ($J_0=0$), the model reduces to a paramagnetic spin model (or ideal spin gas) also in the dynamical regime, and $|\chi_{21}(\omega)|$ is expected to be a decreasing function of $\omega$, with a low-frequency plateau value equal to $1/T^3$, in agreement with the static results of Sec.~\ref{sec:static:resp} [see Eq.~\eqref{eq:chi21static:intspins}].
Including interactions with nonzero quenched couplings $J_{i,i+1}$ leads to the emergence of a coherence length, that grows when decreasing temperature, as discussed below.

Heuristically, and as long as the frequency is not too low, the system may be thought of as an ideal gas of superdipoles, as mentioned above, where superdipoles are composed of typically $N_{\rm corr}(T)$ neighboring spins with an essentially frozen disordered structure.
Hence $\chi_{21}(\omega)$ is expected to remain a decreasing function of $\omega$, but now with a higher low-frequency plateau value (see below).
To get the humped shape, one thus needs to take into account the slow dynamics of the coupling constants, on a time scale much larger than the one of the spin dynamics. In this low frequency regime, the third order response thus goes from the high plateau value $\propto N_{\rm corr}(T)/T^3$ down to the ideal gas non-linear response equal to $1/T^3$.

In its simplest version, the superdipole argument assumes that the $N$ spins can be divided into groups of $\nc$ neighboring spins that flip simultaneously, and thus constitute a superdipole (or `superspin'). Interactions between superdipoles are neglected.
To formulate the argument in a quantitative way, it is thus useful to first evaluate the dynamic response of noninteracting spins.

\subsubsection{Dynamic response of noninteracting spins}

For later generalization to the superdipole case, it is convenient to assume that the spins take values $S_i=\pm\mu$, where $\mu$ is the dipolar moment.
The transition rate for spin reversal is given by
\be \label{eq:rate:S:spd}
W(-S_i|S_i) = \frac{\nu_0}{1+e^{2\beta \mu E S_i}}.
\ee
For noninteracting spins, the dynamic response can be evaluated from the study of a single spin $S$ (where $S$ is any of the spins $S_i$).
Due to the absence of (both static and dynamic) correlations between different spins, the equilibrium magnetization correlation $C(t,E)$ defined in Eq.~(\ref{eq:def:correl}) boils down to the single-spin two-time correlation
\be \label{eq:def:correl:singlespin}
C_S(t,E_0) = \la S(t) S(0) \ra_{E_0} - \la S \ra_{E_0}^2\,.
\ee
For the two-state stochastic process defined by the transition rate (\ref{eq:rate:S:spd}), the time-dependent solution of the master equation can be written down explicitly.
The correlation $C_S(t,E)$ is obtained as
\be
C_S(t,E) = \mu^2 \left( 1-\tanh^2(\beta \mu E)\right) e^{-\nu_0 t}.
\ee
One then obtains from Eqs.~(\ref{eq:def:Comega}) and (\ref{eq:FDT})
\be
\chi(\omega,E) = \frac{\mu^2 \beta}{1-i\omega\tau} \, \left( 1-\tanh^2(\beta \mu E)\right),
\ee
with $\tau=\nu_0^{-1}$.
This respectively leads for the linear and cubic responses to
\be
\chi_1(\omega) = \frac{\mu^2 \beta}{1-i\omega\tau}\,, \qquad
\chi_{21}(\omega) = -\frac{\mu^4 \beta^3}{1-i\omega\tau}\,.
\ee
Static results of Sec.~\ref{sec:static:resp} are recovered in the limit
$\omega\to 0$, for $\mu=1$.
The moduli of the linear and cubic responses are decreasing functions of the frequency,
\be
|\chi_1(\omega)| = \frac{\mu^2 \beta}{\sqrt{1+(\omega\tau)^2}}\,, \quad
|\chi_{21}(\omega)| = \frac{\mu^4 \beta^3}{\sqrt{1+(\omega\tau)^2}}\,.
\ee
Expanding more generally $\chi(\omega,E)$ in powers of $E$,
%$=\sum_n \chi_{2n,1}(\omega) E^{2n}$,
\be
\chi(\omega,E) = \sum_{n=0}^{\infty} \chi_{2n,1}(\omega) E^{2n}
\ee
with $\chi_{0,1}\equiv \chi_1$, one finds in the same way that
\be
|\chi_{2n,1}(\omega)| \propto \frac{\mu^{2+2n} \beta^{1+2n}}{\sqrt{1+(\omega\tau)^2}}
\ee
is a decreasing function of $\omega$.

\subsubsection{Dynamic response of noninteracting superdipoles}

In the superdipole picture, one assumes that interactions make spins move coherently as blocks of $\nc$ spins. Yet, each block of spin has a disordered internal structure due to the glassy nature of the system.
One is thus led to consider superdipoles with a dielectric moment 
$\mu \approx \nc^{1/2}\, \mu_0$, with $\mu_0$ the individual dielectric moment.
Since the dynamic response is evaluated as a density with respect to the number of spins (and not of superdipoles), it has to be further normalized by $\nc$. One ends up with
\be
|\chi_{2n,1}^{\mathrm{sd}}(\omega)| \propto \frac{(\sqrt{\nc}\, \mu_0)^{2+2n} \beta^{1+2n}}{\nc \sqrt{1+(\omega\tau)^2}}
\propto \frac{\nc^n}{\sqrt{1+(\omega\tau)^2}}.
\ee
where the superscript 'sd' stands for ''superdipoles''.
The experimentally observed humped shape of the third and fifth order non-linear responses suggests that at very low frequency, $\nc$ should actually be an increasing function of $\omega$, that saturates to a finite value at higher frequencies. This is consistent with the fact that correlations are weak or even absent at equilibrium.
Assuming a slow enough increase of $\nc(\omega)$, the above calculation still approximately applies, and one finds
\be
|\chi_{2n,1}^{\mathrm{sd}}(\omega)| \propto \frac{\nc(\omega)^n}{\sqrt{1+(\omega\tau)^2}}
\ee
which reproduces the typical humped shape of non-linear responses.
Consistently with experiments, one finds that the linear response is independent of $\nc$, and that for large $\nc$, the nonlinear responses become larger when their order $2n+1$ is increased.

The spin models considered in this work offer an interesting opportunity to assess and substantiate the superdipole picture.
On general grounds, the superdipole size $\nc$ is related to a coherence length
$\xi_{\rm c}$ through a scaling relation $\nc \sim \xi_{\rm c}^{d_f}$, where
$d_f \lesssim d$ is the fractal dimension of correlated clusters of spins.
In the following, we determine the coherence length $\xi_{\rm c}$ and compare it with the length $\xi_{\rm hd}$ characterizing dynamical heterogeneities.
We show that the superdipole scenario is qualitatively recovered when
$\xi_{\rm c} \gg \xi_{\rm hd}$, pointing to a two-length scale scenario which requires the presence of interactions. In contrast, the case $\xi_{\rm c} \approx \xi_{\rm hd}$, corresponding to a single length scale, is found in the KCM where no interactions are present, as discussed below.

\subsection{Coherence length}
\label{sec:cooper:length}

A standard method to determine a dynamical length is to look at the spatial correlation of a persistence variable that compares the local configuration of the system at two different times. In the context of a spin model, it is natural to compare the value of the spin $S_i$ at time $t$ with its value at time $t=0$ (assuming that the system is equilibrated at $t=0$).
We thus introduce the local overlap variable $q_i(t)=S_i(t)S_i(0)$ and define the four-point correlation $g_4^q(r,t)$ as the spatial correlation of the two-time local overlap variable $q_i(t)$:
\be
g_4^q(r,t) = \big< \la q_i(t)q_{i+r}(t)\ra_i - \la q_i(t)\ra \la q_{i+r}(t)\ra_i \big>_{\rm tr}\,
\label{eq:g4q}
\ee
where $\la \dots \ra_i$ denotes a spatial average over site $i$,
and $\la \dots \ra_{\rm tr}$ stands for an ensemble average over stochastic trajectories and initial conditions.
The associated four-point susceptibility $\chi_4^q(t)$ then reads
\be \label{eq:def:chi4:int}
\chi_4^q(t) = \sum_r g_4^q(r,t)\,.
\ee
In the present one-dimensional context, a coherence length $\xi_{\rm c}$ quantifying cooperative effects can then be defined by normalizing $\chi_4^q(t)$ by $g_4^q(0,t)$ as
\be \label{eq:def:xi:int}
\xi_{\rm c}(t) = \frac{\chi_4^q(t)}{g_4^q(0,t)}\,.
\ee
Quite importantly, in our spin model with stochastic couplings, the local overlap $q_i(t)$ does not decorrelate after the first spin flip, but keeps a memory of the initial condition until the coupling constants rearrange.
In other words, the local energy landscape acts as a restoring force for the spin, which keeps taking a given preferred value until the local energy landscape rearranges due to the slow evolution of the couplings.

\begin{figure}[t]
%  \centering
  \includegraphics[width=0.48\textwidth]{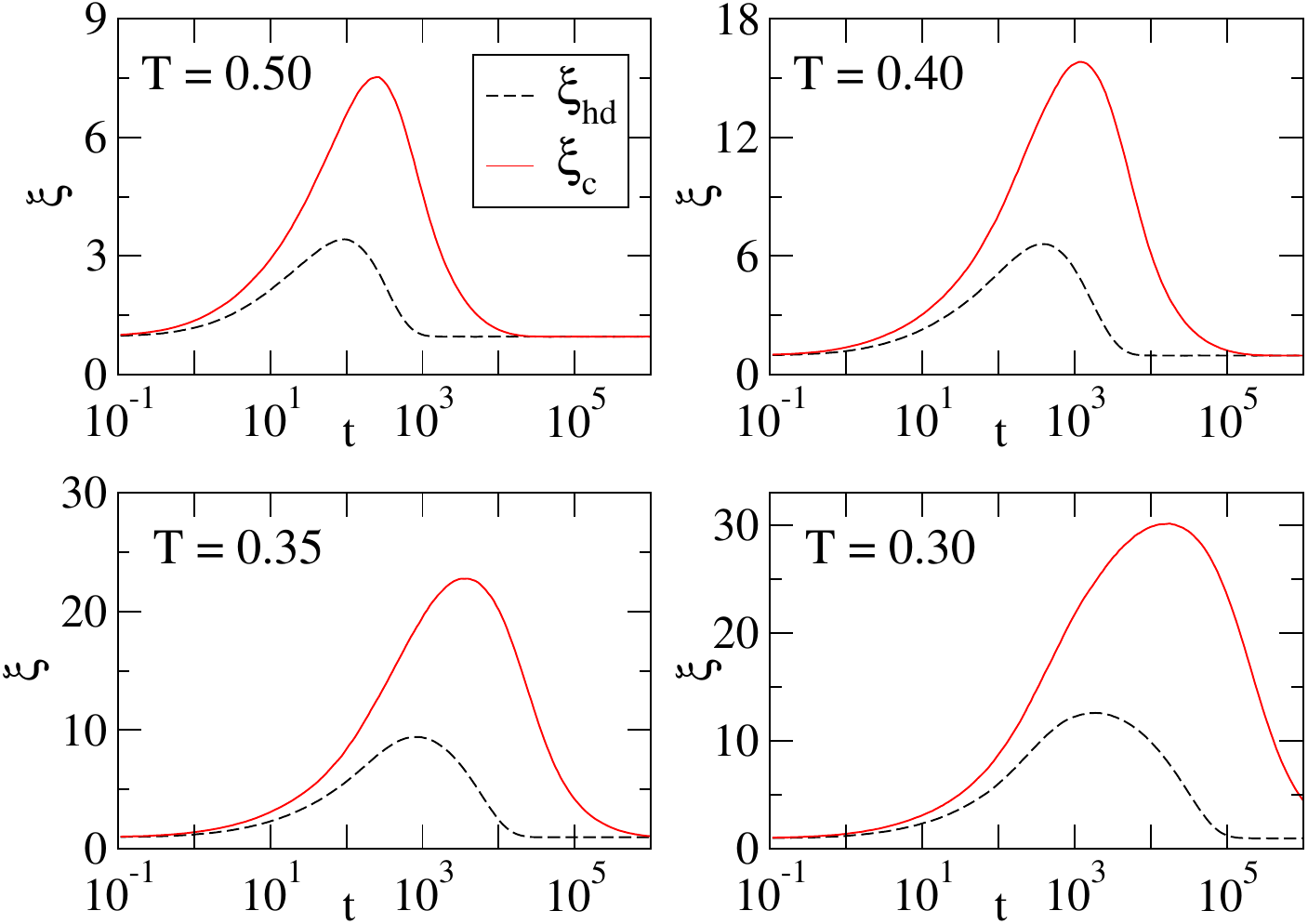}
  
  \caption{Dynamic lengths $\xi_{\rm hd}$ characterizing dynamical heterogeneities (dashed line) and $\xi_{\rm c}$ characterizing cooperative effects (full line) as a function of time $t$ in the spin model with stochastic couplings, for different values of temperature $T$. Note that the $Y$-axis scale changes from one panel to the other to enhance readability. At low temperature, $\xi_{\rm c}(t)$ becomes much larger than $\xi_{\rm hd}(t)$ for $t \gtrsim \tau_{\alpha}$, showing that interactions play the role of a restoring force when couplings change very slowly. Parameters: $J_0=1$, $B=3$, $N=10^3$.}
  \label{fig:chi4interact}
 \end{figure}

\subsection{Dynamical heterogeneities}
\label{sec:HD:length}

We now aim at determining a correlation length $\xi_{\rm hd}$ of dynamical heterogeneities that can be quantitatively compared to the coherence length
$\xi_{\rm c}$. Dynamical heterogeneities in spin models are usually characterized by introducing a local persistence variable $\phi_i(t)$ that satisfies $\phi_i(0)=1$ and keeps the value $\phi_i(t)=1$ as long as the spin $S_i$ does not flip.
A standard choice is then to assign the value $0$ to the persistence variable after the first spin flip, whatever the later spin value
(see, e.g., \cite{Berthier05,Bertin05}).
Here, to remain as close as possible to the overlap variable $q_i(t)$ defined in Sec.~\ref{sec:cooper:length}, we instead assume that at each flip of spin $S_i$, $\phi_i(t)$ is randomly assigned a value $\pm 1$, with equal probability.
In this way, $\phi_i(t)$ takes values $\pm 1$ similarly to $q_i(t)$, but correlations with the value $S_i(0)$ are lost after the first spin flip. In other words, the restoring force is discarded in the definition of $\phi_i(t)$.

We define the four-point correlation function $g_4^{\phi}(r,t)$ as the spatial correlation function of the two-time variables $\phi_i(t)$,
\be
g_4^{\phi}(r,t) = \big< \la \phi_i(t)\phi_{i+r}(t)\ra_i - \la \phi_i(t)\ra \la \phi_{i+r}(t)\ra_i \big>_{\rm tr}
\ee
with the same notations for averages as in Eq.~\eqref{eq:g4q}.
The corresponding four-point susceptibility $\chi_4^{\phi}(t)$ reads as
\be \label{eq:def:chi4:HD}
\chi_4^{\phi}(t) = \sum_r g_4^{\phi}(r,t)\,.
\ee
The correlation length $\xi_{\rm hd}$ characterizing dynamical heterogeneities
is then defined by normalizing $\chi_4^{\phi}(t)$ by $g_4^{\phi}(0,t)$ as
\be \label{eq:def:xi:HD}
\xi_{\rm hd}(t) = \frac{\chi_4^{\phi}(t)}{g_4^{\phi}(0,t)}\,.
\ee

\subsection{Numerical results}
\label{sec:length:num}

\subsubsection{Spin model with stochastic couplings}

We have evaluated numerically the correlation lengths $\xi_{\rm hd}(t)$
and $\xi_{\rm c}(t)$ in the spin model with random couplings defined in Sec.~\ref{sec:spin:interact}. These two length scales are plotted in Fig.~\ref{fig:chi4interact} for different values of temperature $T$.

Lowering temperature, the time scale separation between spin dynamics and coupling dynamics is increased, i.e., $\nu_1(T) / \nu_0 \ll 1$. In this regime, the coherence length $\xi_{\rm c}(t)$ becomes much larger than the characteristic length $\xi_{\rm hd}(t)$ of dynamical heterogeneities for times $t \gtrsim \tau_{\alpha}$, and its maximum shifts to larger times with respect to that of $\xi_{\rm hd}(t)$.
We note in particular that $\xi_{\rm c}(t)$ still takes appreciable values in a time regime when $\xi_{\rm hd}(t)$ has already relaxed to a value close to unity.

\begin{figure}[t]
%  \centering
  \includegraphics[width=0.48\textwidth]{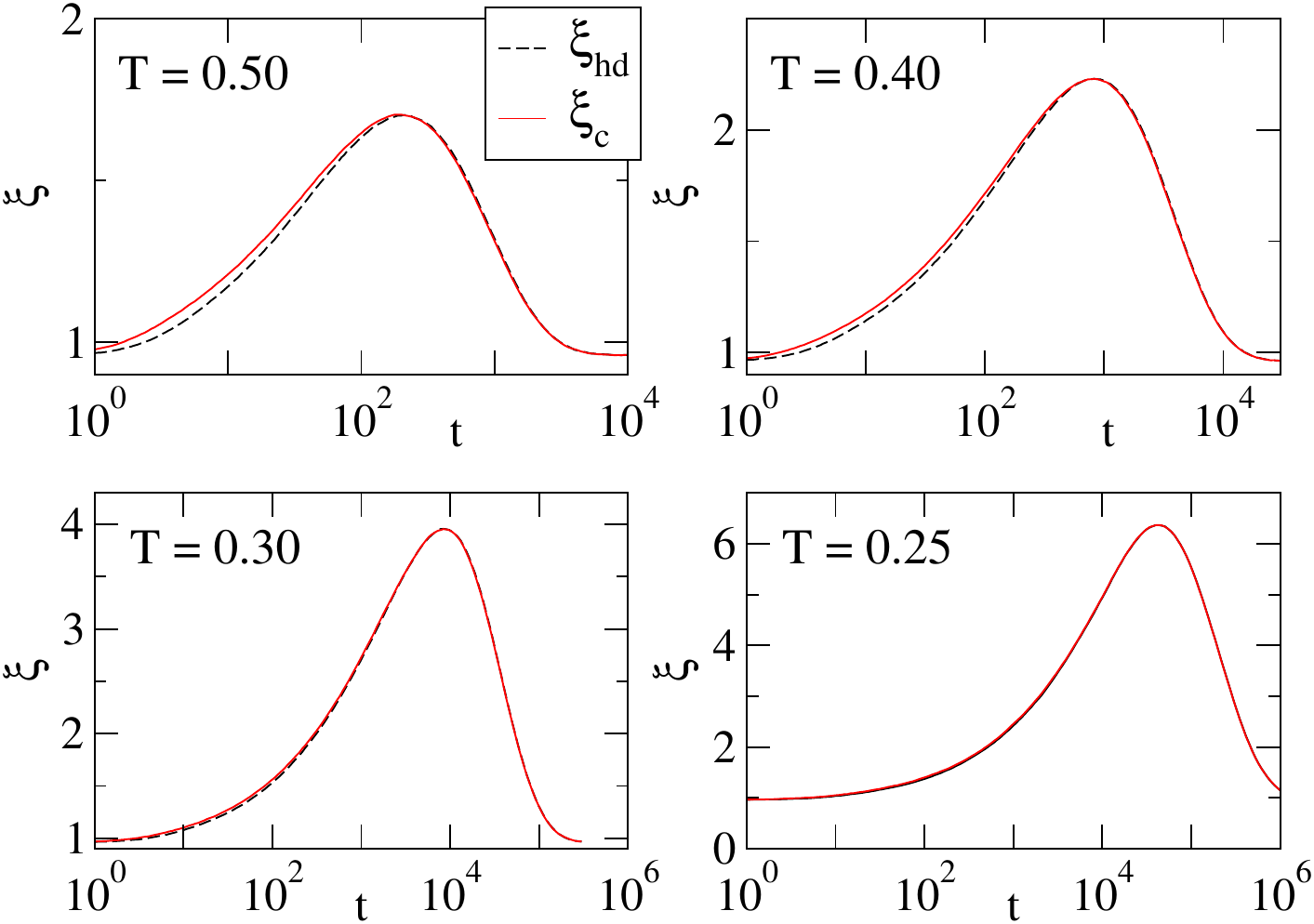}
  \caption{Dynamic lengths $\xi_{\rm hd}$  (dashed line) and $\xi_{\rm c}$ (full line) as a function of time $t$ in the kinetically constrained spin model, for different values of temperature $T$. Parameters: $K=1$, $N=10^3$.}
  \label{fig:chi4kcm}
 \end{figure}

\subsubsection{Kinetically constrained spin model}

To compare these results with the basic facilitation picture, we plot on Fig.~\ref{fig:chi4kcm} the characteristic lengths $\xi_{\rm c}(t)$ and $\xi_{\rm hd}(t)$ for the KCM defined in Sec.~\ref{sec-def-kcm}, for different temperature values. 
We observe that in this case $\xi_{\rm c}$ remains almost identical to $\xi_{\rm hd}$, meaning that glassy relaxation is dominated here by dynamical heterogeneities. This result is consistent with the fact that no landscape-driven restoring force is present in simple kinetically constrained models with no interactions between spins.

%\bigskip

\section{Discussion and conclusion}

We have seen in Sec.~\ref{sec:length:num} that for the model of interacting spins with slowly rearranging couplings, cooperative effects resulting from interactions between spins dominate over purely dynamical heterogeneities in a broad time regime, potentially extending over one decade or more after the relaxation time $\tau_{\alpha}$.
Although cooperative effects are also dynamical here, in the sense that no static spatial correlations are present in the model as discussed in Sec.~\ref{sec:static:resp}, this dynamical cooperativity keeps track of the underlying critical spin-glass physics which only partly unfolds due to the slow rearrangements of the coupling constants.
One might thus speak of an avoided spin-glass transition due to slow rearrangements, which needs two different length scales to be characterized.

It is important to note that no static length is present in the model of interacting spins considered here. Numerically, one finds that both $\xi_{\rm hd}$ and $\xi_{\rm c}$ go to $1$ in the long-time limit. Analytically, one finds that at equilibrium, all degrees of freedom (spins and couplings) are fully decorrelated, see Sec.~\ref{sec:static:resp}. These results on the present simple model of interacting spins are actually consistent with experimental results on the cubic dielectric response, that recover the ideal gas response in the limit of vanishing frequency.

Overall our work illustrates explicitly that the qualitative behavior of nonlinear responses --in temperature and frequency-- changes drastically depending on the presence or absence of interactions between effective degrees of freedom for glass formation. In the framework of the simple spin models studied here, this effect has been traced back to the fact that cubic responses are actually not sensitive to dynamical correlation effects (characterized by $\xi_\mathrm{hd}$) but rather to dynamical coherence effects (characterized by $\xi_\mathrm{c}$). One may expect that the same effect carries over to nonlinear responses of higher order. Physically, such coherence effects only exist if interaction terms in the Hamiltonian explicitly make some spin configurations being preferred with respect to other spin configurations, as long as coupling constants have not yet rearranged --i.e., over a long but finite time scale. In that case only, the system has some finite rigidity, i.e., when perturbed by the field, the system will react as a whole (i.e., a superdipole), unless the field frequency is too small, in which case the superdipoles melt.
Interestingly, recent extensive numerical simulations of glass models at low temperature have shown that dynamic facilitation is at play in the late relaxation stage, for times much larger than the characteristic relaxation time $\tau_{\alpha}$ \cite{Guiselin2022,Scalliet2022}. One may thus wonder whether the eventual melting of superdipoles in our model may involve some type of effective dynamic facilitation, whereby coherent domains could progressively rearrange through, e.g., diffusion of their boundaries. Whether dynamic facilitation is involved or not in this late-stage relaxation, the mere existence of superdipoles for a time window extending significantly beyond $\tau_{\alpha}$ is a clear sign of the key role played by interactions.

While some models with interacting degrees of freedom may be mapped to kinetically constrained models with non-interacting effective degrees of freedom (see, e.g., plaquette models \cite{Newman1999,Garrahan2000,Jack2005a,Jack2005b,Franz2016}), it is important to note that the external field couples to the original, interacting, degrees of freedom --the spins in the plaquette model, rather than the effective plaquette variables. Hence the existence of a mapping to a kinetically constrained model does not imply that the latter suitably describes the non-linear response to a field --unless one couples the field to the original variables, that have to be reexpressed in terms of generally complicated and non-local functions of the effective variables of the kinetically constrained model.

It is of interest to briefly discuss our results in the perspective of existing glass theories. Our finding of an ``avoided spin-glass criticality'' bears some resemblance with the so-called Frustration theory of the glass transition in which geometric frustration prevents criticality to fully unfold \cite{Tarjus2005}. In this scenario there is an avoided critical point $T^\star$ --with $T^\star > T_g$-- around which the order develops only to some finite range. This yields an ever-flowing --though with highly non trivial correlations-- liquid state, and thus an ideal-gas response at zero frequency for nonlinear cubic responses. Another possibility with which our findings are naturally compatible is the unreachable critical point of Random First Order Transition theory inspired from $p$-spin models \cite{Kirkpatrick1989,Wolynes2012}. According to the RFOT scenario, the static coherence length scale is the point-to-set length $\xi_\mathrm{PTS}$ --see Ref. \cite{Biroli2008}--, which diverges at the Kauzmann temperature $T_\mathrm{K}$ --where $T_\mathrm{K} < T_g$. This yields, as well, an ideal gas response at zero frequency for nonlinear cubic response, as anticipated in Ref.~\cite{Bouchaud2005}, because this length scale does not couple directly to a spatially homogeneous external field, as the one used in dielectric experiments \cite{Albert2019,Biroli2021}. 

More generally, one may imagine other unknown scenarios that would be compatible both with experimental measurements of nonlinear responses and, at a qualitative level, with the results we obtained on simple spin models.  These scenarios should be such that: \textit{i)} interactions between degrees of freedom play a major role, favoring some configurations which are not spatially periodic, and which are driven by a critical point which cannot be crossed at equilibrium on human time scales; \textit{ii)} there is an ideal-gas response at any order in the applied field at zero frequency, while, at finite frequencies, qualitative differences arise between linear and nonlinear responses. Denoting by `molecular amorphous ordering' any scenario fulfilling points \textit{i)-ii)}, one of the outcomes of this work is thus to better illustrate what the nonlinear experiments teach us: namely the fact that, upon cooling, molecular amorphous ordering develops \cite{Albert2016,Albert2019,Biroli2021}. However, these experiments do \textit{not} allow one to discriminate between some already existing scenarios of molecular amorphous ordering, and it might be that they turn out to be consistent with yet unexplored ones. Thus, we still have to unveil the microscopic mechanism by which the amorphous ordering --and the associated glass transition-- takes place so often in nature.

\end{document}